\documentclass[conference]{IEEEtran}
\IEEEoverridecommandlockouts

\usepackage{algorithm}
\usepackage{algorithmic}
\usepackage{relsize}
\usepackage{cite}
\usepackage{amsmath,amssymb,amsfonts}
\usepackage{algorithmic}
\usepackage{graphicx}
\usepackage{hyperref}
\usepackage{microtype}
\usepackage{amsmath}
\usepackage{textcomp}
\usepackage[english]{babel}
\usepackage{cleveref}
\usepackage{booktabs}
\usepackage{graphicx}
\usepackage{mathtools}
\usepackage{amsthm}
\usepackage{enumerate}
\usepackage{csquotes}
\usepackage{xcolor}
\def\BibTeX{{\rm B\kern-.05em{\sc i\kern-.025em b}\kern-.08em
    T\kern-.1667em\lower.7ex\hbox{E}\kern-.125emX}}

\usepackage{csquotes}

\DeclareMathOperator{\Ima}{Im}
\DeclareMathOperator{\argmin}{argmin}
\newcommand{\subheading}[1]{\smallskip\noindent\textbf{#1.} }
\newcommand{\R}{\mathbb{R}}
\newcommand{\N}{\mathbb{N}}
\newcommand{\SC}{\textsmaller{SC}}
\newcommand{\mSC}{\mathcal{S}}
\newcommand\acronymcap[1]{\textsmaller{#1}}
\newcommand\acronymmin[1]{#1}
\newcommand\titlecaps[1]{\textsmaller{#1}}
\newcommand\harm{\mathbf{H}}
\newcommand\cs{\mathtt{cluster\_score}}

\newcommand\michael[1]{\noindent{\textcolor{magenta}{[MTS: #1]}}}
\newcommand\vincent[1]{\noindent{\textcolor{magenta}{[VPG: #1]}}}
\newcommand\florian[1]{\noindent{\textcolor{magenta}{[FF: #1]}}}
\newcommand\josef[1]{\noindent{\textcolor{green}{[JH: #1]}}}
\renewcommand\vincent[1]{}\renewcommand\michael[1]{}\renewcommand\josef[1]{}\renewcommand\florian[1]{} 

\begin{document}

\title{Topological Trajectory Classification and Landmark Inference on Simplicial Complexes
\thanks{Funded by the Ministry of Culture and Science (\acronymcap{MKW}) of the German State of North Rhine-Westphalia (``\acronymcap{NRW}-Rückkehrprogramm''), the German Research Council (\acronymcap{DFG}) within Research Training Group \oldstylenums{2236} (\acronymcap{U}\acronymmin{n}\acronymcap{RAV}\acronymmin{e}\acronymcap{L}) and the European Union (\acronymcap{ERC}, \acronymcap{HIGH-HOP}\acronymmin{e}\acronymcap{S}, \oldstylenums{101039827}). Views and opinions expressed are however those of the authors only and do not necessarily reflect those of the European Union or the European Research Council Executive Agency. Neither the European Union nor the granting authority can be held responsible for them.}
}
\author{\IEEEauthorblockN{Vincent P. Grande}
\IEEEauthorblockA{\textit{Dpt.\ of Computer Science} \\
\textit{\titlecaps{RWTH} Aachen University}\\
Aachen, Germany\\
grande@cs.rwth-aachen.de}
\and
\IEEEauthorblockN{Josef Hoppe}
\IEEEauthorblockA{\textit{Dpt.\ of Computer Science} \\
	\textit{\titlecaps{RWTH} Aachen University}\\
	Aachen, Germany\\
	hoppe@cs.rwth-aachen.de}\\
\and
\IEEEauthorblockN{Florian Frantzen}
\IEEEauthorblockA{\textit{Dpt.\ of Computer Science} \\
	\textit{\titlecaps{RWTH} Aachen University}\\
	Aachen, Germany\\
	florian.frantzen@cs.rwth-aachen.de}
\and
\IEEEauthorblockN{Michael T. Schaub}
\IEEEauthorblockA{\textit{Dpt.\ of Computer Science} \\
	\textit{\titlecaps{RWTH} Aachen University}\\
	Aachen, Germany\\
	schaub@cs.rwth-aachen.de}
}

\maketitle

\begin{abstract}
We consider the problem of classifying trajectories on a discrete or discretised $2$-dimensional manifold modelled by a simplicial complex.
Previous works have proposed to project the trajectories into the harmonic eigenspace of the Hodge~Laplacian, and then cluster the resulting embeddings. However, if the considered space has vanishing homology (i.e., no \enquote{holes}), then the harmonic space of the $1$-Hodge~Laplacian is trivial and thus the approach fails. Here we propose to view this issue akin to a sensor placement problem and present an algorithm that aims to learn \enquote{optimal holes} to distinguish a set of given trajectory classes. Specifically, given a set of labelled trajectories, which we interpret as edge-flows on the underlying simplicial complex, we search for $2$-simplices whose deletion results in an optimal separation of the trajectory labels according to the corresponding spectral embedding of the trajectories into the harmonic space.
Finally, we generalise this approach to the unsupervised setting.
\end{abstract}

\section{Introduction}
Trajectory data appears in a wide range of applications and settings.
Human mobility data can be abstracted as trajectories \cite{gonzalez2008understanding}, ocean currents can be studied using trajectories of so-called ocean drifters \cite{sykulski2016lagrangian}, animal movement has attracted the interest of biologists and ecologists \cite{cleasby2019using}, and finally traffic patterns are relevant for city and traffic planning \cite{li2020trajectory}.

These cases do not only differ in the field of application, but also in the domain of the trajectories.
Usually, trajectories are either assumed to live in continuous real dimensional space $\R^d$, or on discretised spaces modelled as graphs.
Simplicial Complexes (SCs) are generalisations of graphs that also model higher-order geometry and topology and have been shown to be useful for trajectory classification using 'landmark' faces \cite{ghosh2018topological,frantzen2021outlier,yin2015decentralized}.
Common to all of these approaches is that they classify trajectories classes based on the relation between the trajectories and the landmarks, in particular, how much each trajectory goes around the landmarks. 
In real-world data, landmarks can correspond to islands separating ocean flows, high mountains disrupting animal movement patterns, or poorly accessible regions of a city dividing traffic flows.
While \cite{ghosh2018topological} use random walks and transformations to obtain edge weights,
\cite{yin2015decentralized} leverage ideas from differential geometry and topological data analysis,
and \cite{frantzen2021outlier} view the problem from a network-science perspective.
From a mathematical point of view, they interpret the trajectories as edge flow data and then project them into the harmonic space, which is defined as the kernel of the so-called Hodge~Laplacian.
All of these approaches rely on prior knowledge of the underlying topology of the network.
However, in many abstract networks and applications, the set of landmarks or the topology is something we \emph{do not know a-priori} and what we need to \emph{infer} from the trajectory data.
This is in general a very complex problem and the \emph{key challenge} that we aim to address in this paper.

\subheading{Related work}
The field of signal processing over simplicial complexes has been studied in \cite{barbarossa2020topological, Schaub:2021, robinson2014topological}, utilizing the Hodge Laplacian and its eigenvectors (see also \cite{grande2024disentangling}).
The flow embedding of trajectories into the harmonic space of the Hodge Laplacian was first considered in \cite{schaub2020random} and used for outlier detection in \cite{frantzen2021outlier}.
Recent literature on trajectory prediction and analysis on topological spaces includes \cite{roddenberry2021principled,ghosh2018topological}.
The opposite problem is considered in \cite{hoppe2024representing, gurugubelli2024simplicial}, where the authors start with a graph \emph{without} higher-order structure and learn simplices/cells to \emph{add} to provide a sparse representation of given edge flows.
Other applications of the harmonic space of the Hodge~Laplacian can be found in \cite{Chen2021,grande2024topf}.

\subheading{Contributions and Outline}
We propose a novel method to classify trajectories on higher-order networks using discrete Hodge Theory.
In particular, our method \textbf{a)} \emph{extracts} the \emph{hidden structure/topology} of the underlying network based on the \emph{trajectory data} using landmarks which serve as \emph{\enquote{holes}},
\textbf{b)} \emph{classifies} the trajectories using the landmarks extracted before, and
\textbf{c)} works both in an \emph{supervised} and an \emph{unsupervised} setting.

We will give background on \SC{}s and discrete Hodge Theory in \Cref{sec:background}, introduce our landmark inference method in \Cref{sec:theory}, will report results on synthetic and real-world data in \Cref{sec:experiments}, and discuss future work in 
\Cref{sec:discussion}.
\section{Background}

\label{sec:background}
In this section, we will give a brief background on simplicial complexes (\SC{}s) and discrete Hodge Theory for processing edge data. For an introduction to algebraic topology or signal processing on \SC{}s, we refer the reader to \cite{Hatcher:2002} and \cite{Schaub:2021}.

\subheading{Simplicial complexes}
Simplicial complexes are a generalisation of graphs that allow for interactions between more than $2$ nodes in a structured way.
A \emph{simplicial complex} $\mSC$ is a pair $(V,S)$ of a finite set $V$ called \emph{vertices} and a set $S$ of non-empty subsets of $V$ such that \emph{(i)} $S$ is closed under taking non-empty subsets and \emph{(ii)} every vertex is contained in a simplex, i.e.\ $\bigcup_{\sigma \in S}\sigma=V$.
We call the $(k+1)$-element subsets of $S$ the \emph{$k$-simplices} and denote them by $S_k$.
As \SC{}s are generalised graphs, we can identify $0$-simplices with nodes, $1$-simplices with edges, and $2$-simplices with \enquote{closed triangles}.

For bookkeeping purposes, we fix an ordering of $V$ which allows us to associate a unique ordered tuple $(v_0,\dots,v_k)$ to every $k$-simplex.
We can view this as a reference orientation on the $k$-simplices for $k>0$.

Furthermore, we denote by the $k$-th signal space $C_k$ the $\R$-vector space $\R[S_k]$ generated by the set of $k$-simplices $S_k$ and set $C_{-1}$ to be trivial.
Elements of $C_1$ are signals on oriented edges and we will thus view them as \emph{edge flows}.
We can then define the \emph{boundary operators} $B_k(\mSC)\colon C_k\to C_{k-1}$ for $k\geq 0$, by  $B_k(\mSC)\colon (v_0,\dots,v_k) \mapsto\sum_{i=0}^k(-1)^i (v_0,\dots,\hat{v_i},\dots,v_k)$, where $\hat{v_i}$ denotes the omission of the $i$-th entry $v_i$.
We will drop $\mSC$ from the notation if the \SC{} is clear from context.
The first boundary operator $B_1$ corresponds to the edge-vertex incidence matrix of the underlying graph of the \SC{}.

\subheading{Discrete Hodge Theory and the Hodge~Laplacian}
As the signal spaces $C_i$ are finite-dimensional and have an inner product induced by the orthonormal simplex basis, the boundary operators $B_k$ have a matrix representation and a  well-defined adjoint $B_k^*$.
This allows us to define the $k$-th \emph{Hodge Laplacian} $L_k$ as $L_k\coloneqq B_k^*B_k+B_{k+1}B_{k+1}^*$, which recovers the notion of the graph Laplacian for $k=0$.
The Hodge Laplacian gives rise to the Hodge decomposition \cite{eckmann1944harmonische, Lim:2020, Schaub:2021}:
\begin{equation}
	\label{eq:HodgeDecomposition}
C_k = \Ima B_{k+1}\oplus \Ima B_k^*\oplus \ker L_k.
\end{equation}
For $k=1$, $\Ima B_{k+1}$ is the space of curl flows, $\Ima B_k^*$ is the space of gradient flows, and $\ker L_k$ the space of harmonic flows, which are each spanned by eigenvectors of $L_k$.
In this paper, we will have a particular interest in the space of \emph{harmonic flows}, as these are intricately connected to the \emph{topology of the underlying network} and represent smooth $\ell_2$-minimal flows around the holes of the network.
This is reflected in the relation $\dim \ker L_k=\beta_k$, asserting that the dimension of $\ker L_k$ corresponds to the Betti number $\beta_k$ counting the number of $d$-dimensional holes in the network \cite{eckmann1944harmonische, friedman1996computing}.
\section{Method}
\begin{figure}[tb!]
	\begin{center}
		\vspace{-0.1in}
		\includegraphics[width=\linewidth]{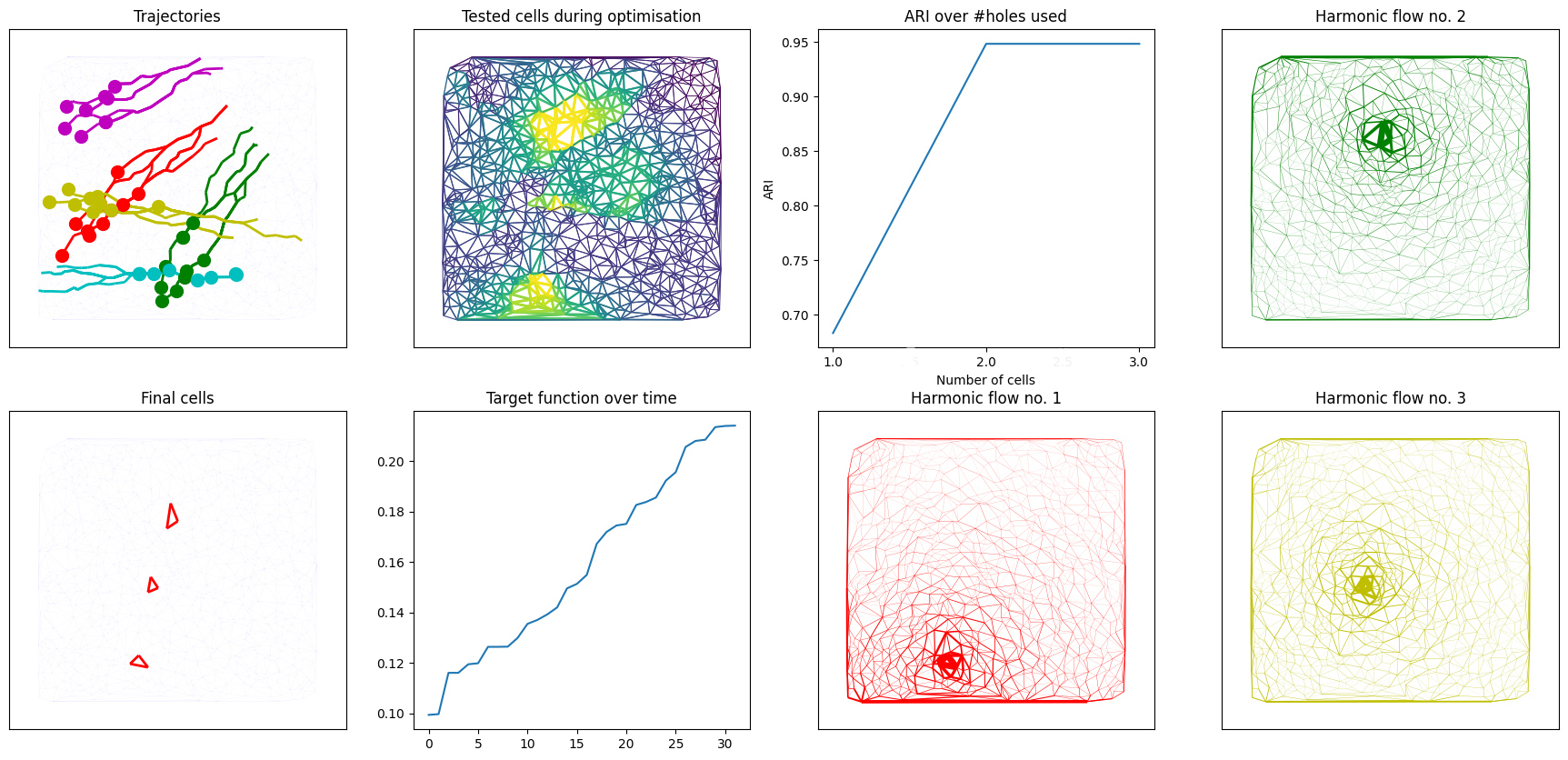}
		\vspace{-0.2in}
		\caption{\textbf{Sample run of the proposed method on a simplicial complex with $3$ holes and $5$ trajectory classes.}
				\emph{Top row, left:} Sample trajectories in the training set, where large dots represent the start of the trajectories.
				\emph{Centre left:} Goal function projected down to simplices, highest possible value shown. Yellow values indicate higher score.
				We can see that the regions of good cluster scores represent the areas between trajectory classes, being good candidates for landmarks chosen by the method.
				\emph{Centre right:} Achieved \textsmaller{ARI} depending on number of holes used for classification. We see that $2$ holes already yield optimal classification performance in this run.
				\emph{Right:} Harmonic flow corresponding to the top chosen $2$-simplex/hole. The strength of the harmonic flow is strongest at the boundary of the hole and then decreases with increasing distance.
				\emph{Bottom left:} Landmarks selected by algorithm, corresponding to simplices which result in good cluster score.
				\emph{Centre left:} Increase of the target function throughout optimisation run.
				\emph{Centre right and right:} Harmonic flows corresponding to remaining two selected landmarks/holes.
		}
		\label{fig:overview}
	\end{center}
	\vspace{-0.3in}
\end{figure}
In this section, we will describe our proposed approach for trajectory classification on simplicial complexes using landmark inference.
For a pseudocode version and an overview of the algorithm, see \Cref{alg:pseudocode} and \Cref{fig:overview} respectively.

\subheading{Problem setting}
We assume that we are given a $2$-dimensional \SC{} $\mSC$ which we will identify with its set $S$ of simplices.
A good intuition for the rest of the theory part will be the case where $\mSC$ arose as a triangulation of a given continuous $2D$ space, but our method works on general $2$-dimensional \SC{}s.
We now assume that a trajectory $t$ in the space of trajectories $T$ is given to us in the form of a sequence of vertices, $t=(v_0,\dots, v_n)$ for some $n$, where adjacent vertices are connected by an edge, i.e.\ $v_i\in \mSC_0$ and $\{v_i,v_{i+1}\}\in \mSC_1$ for all $i$.
The \emph{flow embedding} $F\colon T\to C_1$ assigns an orientation-aligned flow of $1$ to edges along the trajectory and, for the order $<$ on $V$, is given as 
\begin{equation}
	\label{eq:flowembedding}
F((v_0,\dots,v_n))\coloneqq \sum_{\substack{0\leq i<n\\v_i<v_{i+1}}}(v_i,v_{i+1})-\sum_{\substack{0\leq i<n\\v_{i+1}<v_{i}}}(v_{i+1},v_{i}).
\end{equation}
\vspace{-0.2in}

\subheading{Problem formulation}
While the edge flow space $C_1$ captures all trajectory data, it is very high-dimensional and does not behave nicely with respect to the topology and geometry of the underlying space:
Two trajectories $t_1$ and $t_2$ which do not share an edge will get mapped to orthogonal vectors in $C_1$ independent of whether they are right next to each other, or very far apart in the simplicial complex.
We thus need to find a subspace of $C_1$ that \textbf{a)} is low-dimensional, \textbf{b)} maps close trajectories to similar vectors, and \textbf{c)} reflects network topology/geometry.
The \emph{harmonic space} (\Cref{eq:HodgeDecomposition}) is a subspace of $C_1$ that fulfils the above conditions.
Given holes $H_1$ to $H_k$, the associated $0$-eigenvectors $h_1,\dots,h_k$ of $L_1$ represent $\ell_2$-minimal flows that go around their associated holes.
Thus, the projection of $F(t)$ to the subspace spanned by $\harm \coloneqq [h_1\dots h_k]$ is a $k$-dimensional vector storing, for each hole $H_i$, how much $t$ loops around $H_i$.
In particular, for a single hole $H_1$ and a trajectory $t$ that loops around $H_1$ twice, the projection would be $\pm 2$, depending on the orientation of $h_1$ and the direction $t$ loops around $H_1$.

\begin{figure}[t]
	\vspace{-0.15in}
	\begin{algorithm}[H]
		\caption{Proposed topology inference and trajectory classification method.}
		\label{alg:pseudocode}
		\begin{algorithmic}
			\STATE {\bfseries Input:} Simplicial complex $\mSC$, (labelled) trajectories $T_\text{train}$, ($Y_\text{train}$), test trajectories $T_\text{test}$, $n_\text{holes}$, $n_\text{init}$.
			\STATE Compute (diffused) flow embedding $F(T_\text{train})$, \Cref{eq:flowembedding,eq:diffusion}.
			\FOR{$i=1,\dots,n_\text{holes}$}
			\FOR{$j=1,\dots,n_\text{init}$}
			\STATE Pick $2$-cell $\sigma_2^{i,j}\in \mSC_2$ uniformly at random.
			\STATE Compute harmonic vector $h'_i$ corresponding to $\sigma_2^{i,j}$ and harmonic projections ${h'_i}^\top F(T_\text{train})$, \Cref{eq:least_squares}.
			\STATE $c'\gets\text{cluster\_score}\left([h'_1,\dots,h_i]^\top F(T_\text{train}),Y_\text{train}\right)$ \ref{eq:cluster_score},\ref{eq:cluster_score_unsupervised}
			
			\IF{$j=1$ \OR $c'>c$}
			\STATE Set $c\gets c'$,
			$\sigma_2^{i}\gets \sigma_2^{i,j}$,
			$h_i\gets h'_i$.
			\ENDIF
			\ENDFOR
			\ENDFOR
			\FOR{$\Sigma'\in N(\sigma_2^1,\dots,\sigma_2^{n_\text{holes}})$ (\Cref{eq:neighbours})} 
			\STATE Compute harmonic vectors $\mathbf{H}'$ of $\Sigma'$, (\Cref{eq:least_squares}).
			\STATE Set $c'\gets \text{cluster\_score}(\mathbf{H}'^\top F(T_\text{train}))$, (\ref{eq:cluster_score}, \ref{eq:cluster_score_unsupervised}).
			\IF{$c'>c$}
			\STATE Set $(\sigma_2^1,\dots,\sigma_2^{n_\text{holes}})\gets \Sigma'$, $c\gets c'$, $\mathbf{H}\gets\mathbf{H'}$
			\STATE \textbf{restart for loop} (with new neighborhood)
			\ENDIF
			\ENDFOR
			\STATE Train random-forest classifier $\mathcal{X}$ on $\mathbf{H}^\top F(T_\text{train}),Y_\text{train}$.
			\STATE Classify labels $Y_\text{test}\gets\mathcal{X}(\mathbf{H}^\top F(T_\text{test}))$.
			\STATE \textbf{Output:} Landmarks $\sigma_2^1,\dots, \sigma_2^{n_\text{holes}}$, labels $Y_\text{test}$, $\mathbf{H}$, $\mathcal{X}$.
		\end{algorithmic}
	\end{algorithm}
	\vspace{-0.3in}
\end{figure}

Thus, given a finite sequence of trajectories $T_\text{train} = [t_1,t_2,\dots ]$ and associated labels $Y_\text{train}=(y_1, y_2, \dots)$ and a number of holes $k\in\N$, we aim to find the set of holes $H_1,\dots,H_k$ with associated harmonic vectors $\harm(\mSC\setminus H_1,\dots,H_k)$ that optimise some cluster score:
\begin{equation}
	\max_{\substack{  H_i\subset \mSC_2\\\text{for }1\leq i\leq k\\
		H_i\text{ conn.}}} \cs\left(\harm(\mSC\setminus H_1,\dots H_k) F (T_\text{train}), Y_\text{train}\right).
\end{equation}
Here, $\cs(X,Y)$ is a measure like 
\begin{equation}
\label{eq:cluster_score}
\cs(X,Y)=\frac{\min_{i,j:y_i\neq y_j}\lVert x_i-x_j\rVert_2}
{\max_{i\neq j:y_i=y_j}\lVert x_i-x_j\rVert_2}.
\end{equation}
We will then assess the performance of the method by the clustering accuracies using a random-forest or a $k$-nearest neighbours classifier on the training data and the projection of $T_\text{test}$ into the harmonic space.

The above optimisation problem has a large and complicated solution space due to the number and arbitrary shape of holes.
In practice, this makes it infeasible to solve exactly for any meaningful input size.
We thus propose a solver that considers a more constrained solution space and uses approximations to further reduce computational load.
In particular, we will \textbf{1)} assume that all holes $H_i$ consists of a single removed $2$-simplex $\sigma_2^i$.
This greatly reduces the size and complexity of the search space, while providing a good approximation to the behaviour of arbitrary holes.
Furthermore, \textbf{2)} instead of computing the harmonic vectors associated to $H_i$ using the simplicial complex $\mSC\setminus H_1,\dots, H_k$ with all holes removed, we will use the simplicial complex $\mSC\setminus H_i$.
This allows us to not recompute the harmonic embeddings wrt.\ $H_i$ when changing $H_{j}$ for $i\not =j$, greatly reducing the computational burden. Unless holes $i$ and $j$ are very close to each other, this is a good approximation of the harmonic vectors.

\subheading{Computing the Harmonic Eigenvectors}
The authors of \cite{frantzen2021outlier} computed the harmonic space $\harm$ using an eigenvector solver for the Hodge~Laplacian $L_1$.
However, eigenvector computations for large matrices are expensive, and do not necessarily return an eigenvector basis corresponding to the individual holes.
Instead, the boundary $B_2\sigma_2^i$ of the $2$-simplex $\sigma_2^i$ corresponding to hole $H_i$ is a gradient-free (because $B_1B_2=0$) flow around hole $H_i$.
Thus, $B_2\sigma_2^i$ consists of a curl and the desired harmonic part.
We can obtain the curl part by computing
\begin{equation}
	(B_2(\mSC)\sigma_2^i)_\text{curl}=B_2(\mSC\setminus H_i)B^\dagger_2(\mSC\setminus H_i)B_2(\mSC)\sigma_2^i
\end{equation}
where we do not need to compute the Moore--Penrose pseudoinverse explicitly, but can rather compute the computationally efficient sparse least-squares problem
\begin{equation}
	\label{eq:least_squares}
B^\dagger_2(\mSC\setminus H_i)B_2(\mSC)\sigma_2^i = \argmin_x \lVert B_2(\mSC)\sigma_2^i -B_2(\mSC\setminus H_i)x\rVert_2^2.
\end{equation}
We then obtain the desired harmonic eigenvector $h_i$ by normalising $B_2\sigma_2^i-(B_2\sigma_2^i)_\text{curl}$.
The above computation is efficient even for very large sparse \SC{}s, and thus allows us to compute the harmonic eigenvectors of a large number of different holes.

\subheading{Exploring the search space}
Although we have now established means to compute the harmonic projections efficiently, we still need a way to tackle the complex search space of $k$-tuples of $2$-simplices $\mSC_2^k$.
We will do this by some form of local search: given a $k$-tuple of $2$-simplices $\Sigma = (\sigma_2^1,\dots,\sigma_2^k)$, we will denote their set of neighbours in search space
\begin{equation}
	\label{eq:neighbours}
\left\{(\sigma_2^1,\dots,\tilde{\sigma}_2^i,\dots,\sigma_2^k) : 1\leq i\leq k, \lvert\sigma_2^i\cap \tilde{\sigma}_2^i\rvert=2\right\}
\end{equation}
by $N\left((\sigma_2^1,\dots,\sigma_2^k)\right)$.
For a goal function $c\colon \mSC_2^k\to \R$, we would like to compute $c$ on the current $\Sigma\in \mSC_2^k$ and compare it with $c$ on $N(\Sigma)$ to obtain some form of local gradient to update $\Sigma$.
Due to the dimensionality and size of the search space, in practice we will set $\Sigma\coloneqq\Sigma'$ for the first $\Sigma'\in N(\Sigma)$ with $c(\Sigma')>c(\Sigma)$ while storing both all computed $\Sigma'\in \mSC_2^k$ and all computed harmonic eigenvectors $h(\sigma_2)$, reusing them when needed.
Furthermore, we can increase the step size by allowing $n$-hop neighbouring $2$-simplices for the computation of $N(\Sigma)$, further optimising speed.
The method will then stop when reaching a local optimum within its $n$-hop neighbourhood.

\subheading{Diffusion of Trajectories}
\begin{figure}[tb!]
	\vspace{-0.1in}
	\begin{center}
		\includegraphics[width=0.49\linewidth]{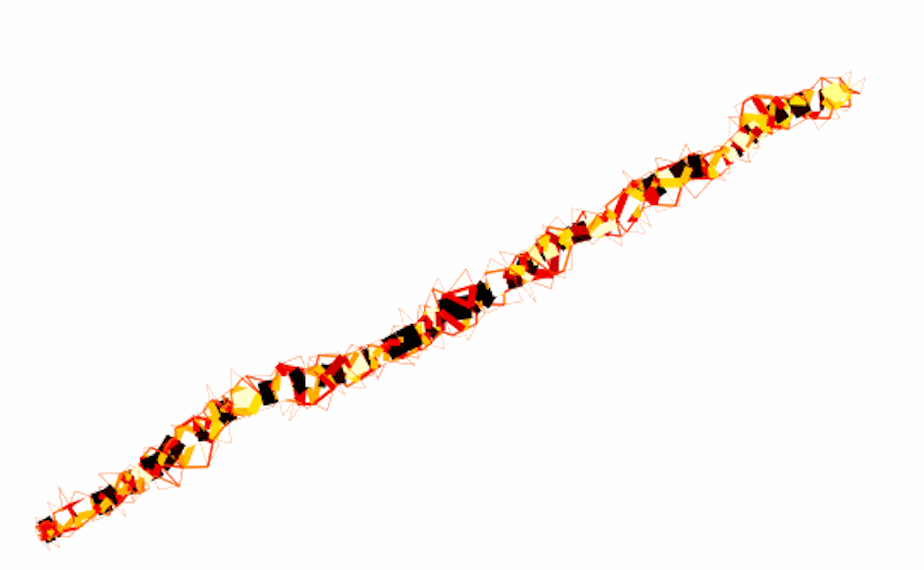}
		\includegraphics[width=0.49\linewidth]{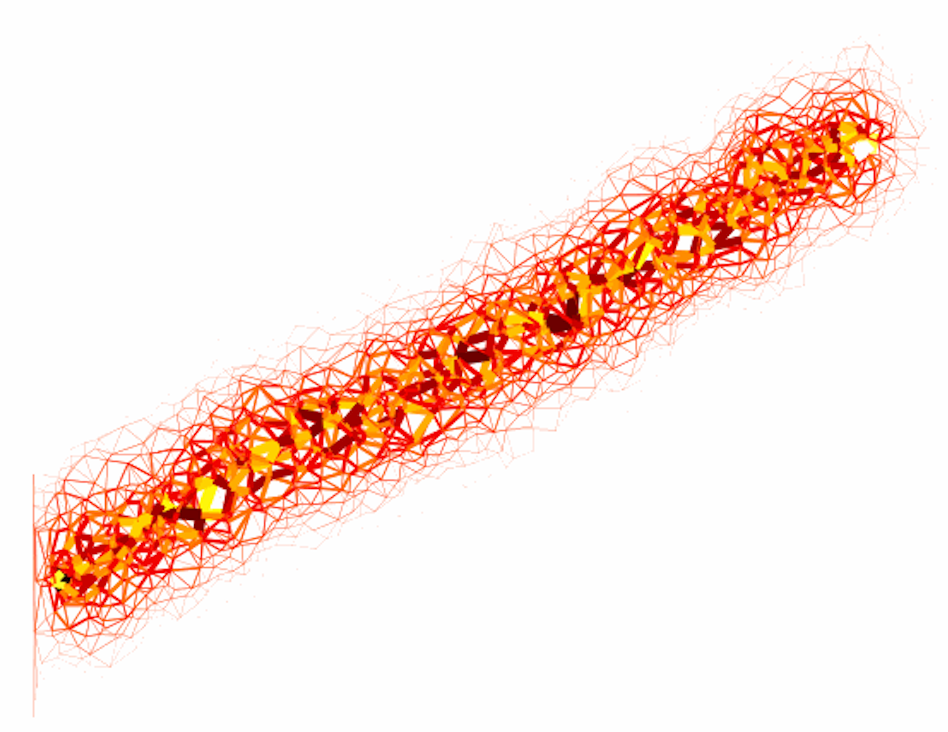}
		\vspace{-0.2in}
		\caption{\textbf{Diffusion processes on edge flows}
			\emph{Left:}
			Trajectory under diffusion for small time step $\tau$.
			\emph{Right:}
			Trajectory under diffusion for larger time step $\tau$.
			We do not show the underlying network.
		}
		\label{fig:diffusion}
	\end{center}
	\vspace{-0.3in}
\end{figure}
After having set up our strategy for the exploration of the search space, we still face the problem that the goal function might behave discontinuously, especially when crossing the trajectories.
This is a major obstacle for any gradient-like optimisation problem.
Thus, we propose to \emph{diffuse the flow embeddings} of the trajectories making them less localised.
This not only counteracts overfitting, but makes the harmonic projections and thus the goal function behave more smoothly with respect to the selected holes.
Just as ordinary diffusion processes on the vertices are governed by the Laplacian via $\dot{x
}=-L_0x$, diffusion processes on oriented edges are governed by the Hodge~Laplacian.
However, in contrast to the $0$-dimensional case, in the $1$-dimensional we have the choice whether to degrade the curl, the gradient, or both classes of eigenvectors in the diffusion process.
As we want to keep the directionality of the trajectories, we opt to only degrade the curl eigenvectors and obtain the differential equation and closed form solution
\begin{align}
	\dot{f}&=-L_1^\text{up}f\\
	\label{eq:diffusion}
	f(T)&=\exp(- L_1^\text{up} \tau) f(0)
\end{align}
  for an edge flow $f$ and time $\tau$, see \Cref{fig:diffusion} for an illustration.
Because $L_1^\text{up}$ is sparse, this product can be efficiently approximated using the sparse matrix exponential.
We will then use the smoothed $\exp(- L_1^\text{up} T) F(t)$ for a $\tau$ depending on the size of the \SC{} as an input to the later steps of the algorithm.

\subheading{Initialisation}
In a naive implementation, we would like to initialise the candidate holes by uniformly sampling the search space $\mSC_2^k$.
However, for large $k$ this becomes infeasible due to high dimensions and the scarcity of \enquote{good} candidate cells.
Instead, we build the initial candidate set \emph{iteratively}:
We first sample $n_\text{init}$ $2$-simplices and choose the simplex $\sigma_2^1$ optimal in the setting with a single hole.
In the next step, we fix $\sigma_2^1$ and sample candidates for $\sigma_2^2$.
This process then continues up to $\sigma_2^k$.
This requires only $kn_\text{init}$ steps instead of the $n_\text{init}^k$ when attempting to evaluate all possible combinations while yielding results of similar quality.
An alternative way would be to start with some configuration and then apply the same search exploration technique as explained above but without the neighbouring condition on the simplices.

\subheading{Unsupervised setting}
When using an unsupervised cluster score function incentivising homogeneous cluster sizes (Eq.\ \ref{eq:cluster_score_unsupervised}), the above method also works in the unsupervised setting.
\label{sec:theory}
\section{Experiments}
\label{sec:experiments}
\subsection{Synthetic Experiments}
\begin{figure}[t!]
	\vspace{-0.1in}
\begin{center}
	\includegraphics[width=0.55\linewidth]{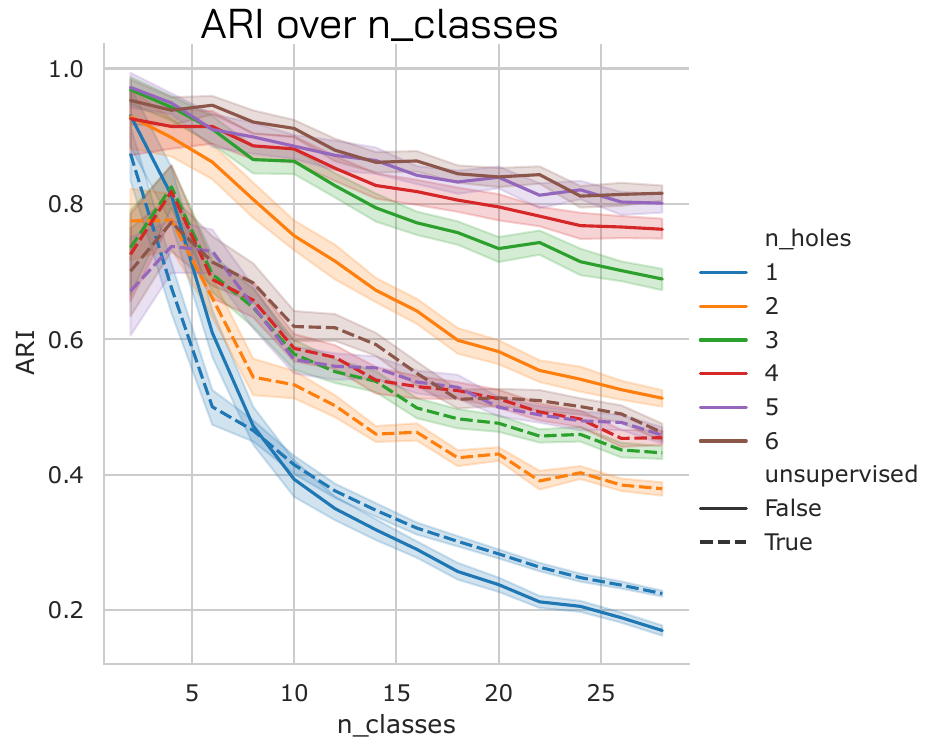}
	\includegraphics[width=0.43\linewidth]{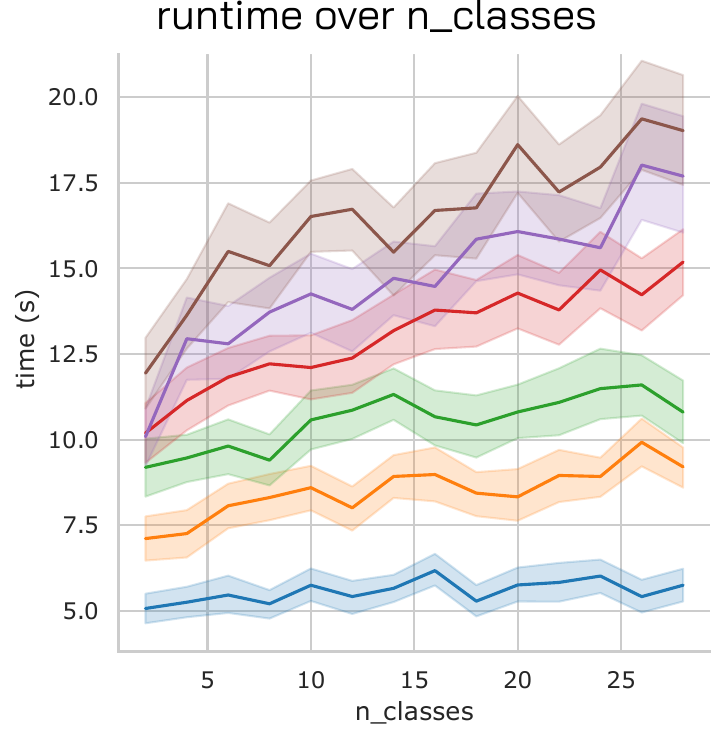}
	\vspace{-0.2in}
	\caption{\textbf{Performance in synthetic experiments.}
		We set $n_\text{train}=5n_\text{classes}$ and $n_\text{test}=50n_\text{classes}$, showing that our proposed method works even for few trajectory examples.
		We vary the number of total classes and report the adjusted rand index (\textsmaller{ARI}, \emph{left}, takes values in $[-0.5,1]$, $1$ for perfect recovery, $0$ for random cluster assignments.) and time (only for the supervised case, \emph{right}).
		Because we keep the size of the network fixed, more classes not only require a more fine grained classifier, but they also are more likely to overlap and thus lead to a harder problem.
		$n_\text{holes}$ indicate the number of holes/landmarks the method is allowed to introduce to classify the trajectories.
		}
	\label{fig:expsynth}
\end{center}
\vspace{-0.3in}
\end{figure}
\label{subsec:synthexperiments}
\subheading{Dataset Creation}
We create synthetic trajectories similarly to \cite{frantzen2021outlier}.
We sample $N$ points randomly in a $[0,1]\times[0,1]$ square and construct a Delaunay triangulation as the \SC{}.
For each trajectory class, we sample a start point $S$ and an end point $E$ near the boundary of the \SC{}.
For the trajectory class, we will then iteratively construct shortest paths between $S$ and $E$, increasing the costs/weights of the used edges after every iteration, which we will then split into test and training set.

\subheading{Supervised results}
In \Cref{fig:expsynth}, we report the performance of the proposed method with increasing number of trajectory classes.
We train the method on $5$ trajectories per example, selecting holes in order to maximise the \texttt{cluster\_score}.
We then train a random-forest classifier on the corresponding harmonic embeddings of the training dataset, which we then use to classify the harmonic embeddings of the trajectories of the test set.
We use the cluster score given in \Cref{eq:cluster_score}.

\subheading{Unsupervised results}
We repeat the setup of the supervised case but do not pass ground truth labels to the algorithm.
We instead run $k$-means clustering on the harmonic embeddings to determine cluster labels.
As a cluster score, we will use a cluster function incentivising clusters of equal size:
\begin{equation}
	\label{eq:cluster_score_unsupervised}
\min_{i,j:y_i\neq y_j}\lVert x_i-x_j\rVert_2	\frac{\min_{k}\left\lvert \{y_i=k : i\le n_\text{classes}\}\right\rvert}
	{\operatorname{std}_k\left\lvert \{ y_i=k : i\le n_\text{classes}\}\right\rvert }
\end{equation}
\subsection{Experiments on Ocean Drifters}
\begin{figure}[t!]
	\begin{center}
		\vspace{-0.1in}
		\includegraphics[width=0.75\linewidth]{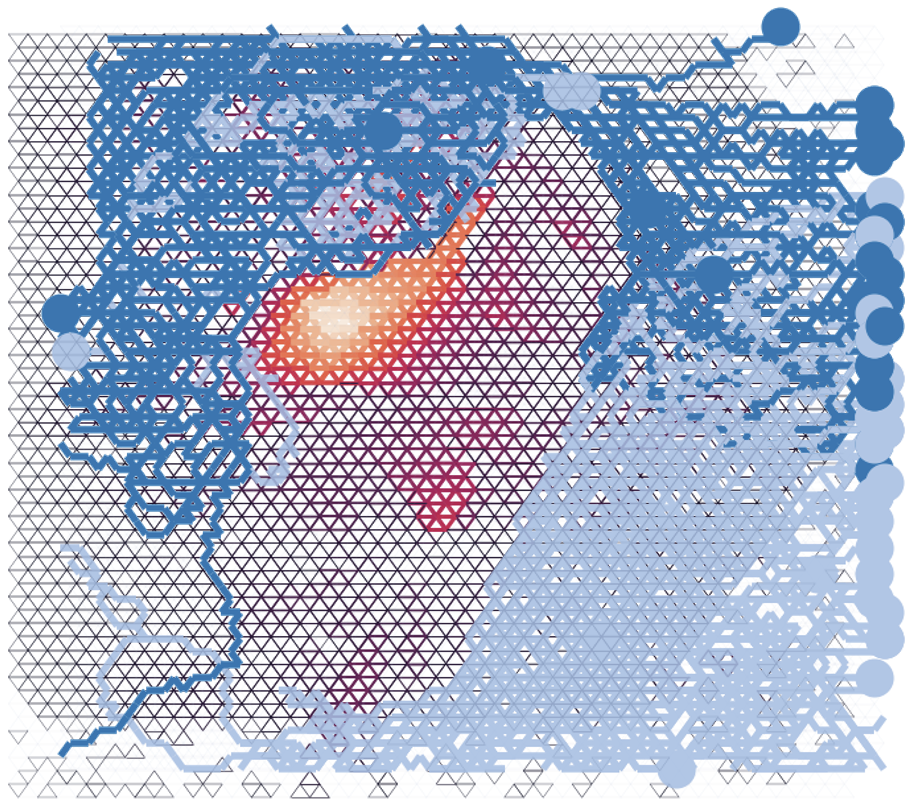}
		\vspace{-0.1in}
		\caption{\textbf{Unsupervised trajectory classification on the Ocean Drifter dataset.}
			We overlay the goal function of the simplices (\emph{red}) with the two trajectory classes (\emph{blue}), and the grid (\emph{black}).
			The method is able to capture a landmark corresponding to the north-west coast of Madagascar.
			Furthermore, it produces two clusters corresponding to drifters passing to the north and to the west of Madagascar, without having access to any labelled data.
		}
			\label{fig:oceandriftermadagascar}
		\end{center}
		\vspace{-0.3in}
	\end{figure}
\label{subsec:drifterexperiments}
In \Cref{fig:oceandriftermadagascar}, we conduct experiments on a dataset from the \href{https://www.aoml.noaa.gov/phod/gdp/data.php}{Global Drifter Program}.
The global drifter program collects data from over 1000 buoys in the oceans of the earth updating their position every 6 hours.
We construct a simplicial complex following the procedure described in \cite{jia2019graph},
where we construct a hexagonal grid on earth's surface, discretise the drifters' trajectories and convert the hexagonal grid into a triangulation corresponding to an \SC.
We then restrict the dataset to the area around the island of Madagascar.
Note that in contrast to \cite{jia2019graph}, we did not remove simplices corresponding to land masses, as we wanted to show that we can learn these landmarks without prior knowledge.
Our method correctly infers a landmark corresponding to the north-west coast of Madagascar, classifying the trajectories based on whether they pass Madagascar in the north or in the south.
\section{Discussion}
\label{sec:discussion}
\subheading{Summary}
We have introduced a method that extracts a \emph{low-dimensional representation} of trajectory data on networks of \emph{arbitrary topology} by \emph{inferring landmarks} and considering the associated \emph{harmonic space} of the Hodge~Laplacian.
This low-dimensional representation can then be used to perform trajectory classification.
The proposed method uses a discrete \emph{local gradient-like optimisation} on the network, employs insights from \emph{Hodge theory} to diffuse the input trajectories to work with \emph{very small $n_\text{train}$ ($<5$)} and \emph{avoid overfitting}, and uses computational tricks to replace eigenvector computations by efficient sparse least-squares problems and can thus be used on \emph{very large networks with $\sim$ 100K nodes} in practice.
By employing a suitable \texttt{cluster\_score}, the method even works in the \emph{unsupervised setting}.
We evaluate the method on synthetic data and real-world data from ocean drifters.

\subheading{Future work}
The proposed method can be directly applied to trajectories on cellular complexes or cubical complexes, as the Hodge Laplacian and the Hodge decomposition exist for all discretised versions of topological spaces.
For future work, it would be interesting to investigate further possible goal functions which either might lead to \emph{direct computation of gradients} or to provable \emph{guarantees on the optimality} of the results found by the heuristic.

Furthermore, taking an orthogonal approach to the proposed method might be an interesting avenue of future research:
Instead of locally searching for the optimal holes we could relax the problem and globally optimise the weights on the $2$-simplices yielding the optimal curl-representatives.
We could then turn the weightings back to landmarks by removing the $2$-simplices with the smallest corresponding weights.
However, this requires ideas overcoming computational difficulties as a straight-forward implementation of the gradient computation on the weight matrix of the above problem involves computing the non-sparse pseudo-inverse matrix of a sparse matrix, which is not feasible for reasonably sized networks.

\subheading{Code}
An implementation of the topological trajectory classification and landmark inference method and an example Jupyter Notebook can be found at \url{https://git.rwth-aachen.de/netsci/landmark-inference}.
\bibliographystyle{IEEEtran}
\bibliography{holes.bib}
\end{document}